\documentstyle[11pt,aaspp4]{article}

\slugcomment{Accepted for ApJ. Lett.}

\lefthead{Kaastra et al.}
\righthead{Nonthermal X-ray emission in A\,2199}

\begin{document}

\title{High- and low energy nonthermal X-ray emission\\
    from the cluster of galaxies A\,2199}

\author{Jelle S. Kaastra}
\affil{Space Research Organization of the Netherlands,
Sorbonnelaan 2, 3584 CA Utrecht, The Netherlands}

\author{Richard Lieu}
\affil{Department of Physics, University of Alabama, Huntsville, AL 35899}

\author{Jonathan P.D. Mittaz}
\affil{Mullard Space Science Laboratory, UCL, Holmbury St.  Mary, Dorking,
Surrey RH5 6NT, England, UK}

\author{Johan A.M. Bleeker and Rolf Mewe}
\affil{Space Research Organization of the Netherlands,
Sorbonnelaan 2, 3584 CA Utrecht, The Netherlands}

\author{Sergio Colafrancesco}
\affil{Osservatorio Astronomico di Roma, Via dell'Osservatorio 2,
I-00040 Monteporzio, Italy}

\and

\author{Felix J. Lockman}
\affil{National Radio Astronomy Observatory, P.O. Box 2, Green Bank, WV 24944}

\begin{abstract}

We report the detection of both soft and hard excess X-ray emission in the
cluster of galaxies A\,2199, based upon spatially resolved spectroscopy with
data from the BeppoSAX, EUVE and ROSAT missions.  The excess emission is visible
at radii larger than 300\,kpc and increases in strength relative to the
isothermal component.  The total 0.1--100\,keV luminosity of this component is
15\,\% of the cluster luminosity, but it dominates the cluster luminosity at
high and low energies.  We argue that the most plausible interpretation of the
excess emission is an inverse Compton interaction between the cosmic microwave
background and relativistic electrons in the cluster.  The observed spatial
distribution of the non-thermal component implies that there is a large halo of
cosmic ray electrons between 0.5--1.5\,Mpc surrounding the cluster core.  The
prominent existence of this component has cosmological implications, as it is
significantly changing our picture of a clusters's particle acceleration
history, dynamics between the thermal and relativistic media, and total mass
budgets.

\end{abstract}


\section{Introduction}

X-ray emission from clusters of galaxies was detected already in the early days
of X-ray astronomy.  Initially the origin of the emission was not clear:
thermal Bremsstrahlung or non-thermal Inverse Compton emission were two common
explanations.  The detection of iron line emission changed the common
interpretation in favour of a thermal origin, although searches for non-thermal
X-ray emission have been continued.

Rather surprisingly, the Extreme UltraViolet Explorer (EUVE) found evidence
for a soft X-ray excess in two of the brightest and most nearby clusters:
Coma (\cite{lieu96a}) and Virgo (\cite{lieu96b}). Later it was also found
in A~1795 (\cite{mittaz98}), and \cite{bowyer98} mention its detection
in A\,2199 and A\,4038. This soft excess is unrelated to the well-known
cooling flow, since it is also observed in Coma (which does not contain
a cooling flow) and its relative strength as compared to the thermal
component increases outwards for three of the five clusters. Detailed
modelling favours a non-thermal origin for the soft excess, for instance
inverse-Compton (IC) emission by cosmic ray electrons on the cosmic microwave
background radiation (see e.g. \cite{sarazin98}).

At the high-energy part of the spectrum, the high sensitivity of the PDS
instrument aboard BeppoSAX allowed the detection of a hard excess in
A\,2199 (\cite{kaastra98}) and Coma (\cite{fusco99}). In the case of
Coma, the data are consistent with the IC interpretation and with the
radio spectrum (\cite{fusco99,lieu99}).

In this paper we analyze the X-ray data from A\,2199, a cluster where both a
soft and hard excess was reported.  A\,2199 is a bright cooling-flow cluster at
redshift 0.0303, with a very low galactic column density of $8.1\times
10^{23}$~m$^{-2}$, as measured accurately by one of us (Lockman).  This allows
the spectrum to be observed down to energies of $\sim$0.1~keV.  We use data
obtained by all narrow-field instruments of BeppoSAX, the DS instrument aboard
of EUVE and the ROSAT PSPC detector.

\section{Data reduction}

For the purpose of our data analysis, the cluster was divided into 7 concentric
annuli, centered around the bright cD galaxy, with outer radii of 3, 6, 9, 12,
15, 18 and 24\arcmin.  The cooling flow is almost completely contained within
the central annulus (0.16\,Mpc).  We use $H_0=50$~km\,s$^{-1}$\,Mpc$^{-1}$ and
$q_0=0.5$ throughout this paper.

The instruments of BeppoSAX are described by \cite{boella97} and references
therein.  Briefly, the LECS (1 instrument) and MECS (3 nearly identical
instruments) are imaging GSPCs with an energy resolution that varies between
0.05--0.7\,keV for energies between 0.1--10\,keV (LECS) and 0.2--0.6\,keV for
energies between 1--10\,keV (MECS).  The spatial resolution is of the order of
an arcminute but degrades towards low energies.  The HPGSPC and PDS
(12--100\,keV) are non-imaging, collimated detectors suited for observing medium
(above 4\,keV) and high-energy (above 12\,keV) X-rays, respectively.

The BeppoSAX observations were obtained between 21--23 April 1997 with an
effective exposure time of 100\,ks for the MECS instrument.  LECS spectra were
extracted for the 6 annuli within 18\arcmin, and MECS spectra for all 7 annuli.  All subsequent data analysis and background subtraction was done
using the January 1999 calibration release of BeppoSAX.

The background for the LECS and MECS spectra was subtracted using the standard
background files taken at empty fields at high galactic latitude.  The
background consists of a cosmic X-ray and a particle contribution.  The total
MECS background during the week of observation was constant,
0.187$\pm$0.004~c/s, and within its statistical uncertainty of 2\,\% consistent
with the long-term (3 months) background (0.191$\pm$0.001~c/s).  This background
is at the same level of the average background of the blank fields of Deep SAX
exposures (0.187~c/s).  The MECS spectra were extracted for the individual units
separately and combined after background subtraction.

The January 1999 SAXDAS release has been used for the effective area,
point-spread function and vignetting of the MECS data.  Strongback obscuration
effects in the MECS have been taken into account by assuming 550\,$\mu$m of Be
with a geometry as described by \cite{boella97}, in addition to the 50\,$\mu$m
that is present for the entire field of view.  Note that near the strongback
around 10\arcmin\ off-axis there may be some systematic uncertainty of at most
20\,\% in the transmission of the strongback, due to lacking calibration data.
The vignetting correction for the LECS was derived from the SAXDAS/LEMAT
raytrace code, assuming azimuthal symmetry around the appropriate center.  The
correction for the support grid was also derived from that package.  The effects
of vignetting and strongback obscuration for the MECS and LECS data have been
taken into account in the response matrices created for this observation.
Collimator vignetting for the HPGSPC and PDS have also been taken into account.
The spectral data for each instrument and region have been rebinned to one third
of the FWHM and in regions with poorer statistics even more.  Contamination by
the nearby AGN EXO\,1627.3+4014 (at 32\arcmin\ from the core of A\,2199) in the
PDS spectrum is unlikely given its steep X-ray spectrum (photon index 3, cf.
Rosat Bright Source Catalog).

The EUVE data were obtained in 1997 (June 16-18), with an effective exposure
time of 49~ks.  Data from the central 3\arcmin\ annulus were omitted from the
analysis since this region contains the dead spot of the DS detector.  We used
data for the 4 annuli between 3--15\arcmin.  Outside that radius the signal is
too weak.

The ROSAT PSPC data were obtained during the PV phase of ROSAT on 18--19 July
1990 with an exposure time of 8.6~ks.  The PSPC data were rebinned into 12 data
channels with a width of about 1/3 of the instrumental FWHM and only data from
channel 47--198 (about 0.39--1.98~keV) were used.  The reason for omitting the
lowest PSPC data channels is that we were not able to obtain spectral fits that
are consistent with the LECS and EUVE data if this energy range is included.
The PSPC calibration for very soft sources is known to be problematic in this
energy range, and moreover the PSPC spectral resolution is twice as poor as the
LECS spectral resolution in the same energy band.  A systematic error of 1~\%
was added to the count rate for all data channels. Rosat spectra were extracted
for the 6 annuli within a radius of 18\arcmin, and also for an additional
annulus between 25--28\arcmin. In between there is too much uncertainty
due to the strongback structure of the PSPC detector. 

The point-spread function (psf) of the LECS and MECS instruments is a sensitive
function of energy.  For example, for the LECS the 50\% encircled energy radius
varies from 4\arcmin\ at 0.2\,keV to 1.2\arcmin\ at 8\,keV.  The core radius and
the size of the cooling flow is about 2\arcmin\ (\cite{jones84,siddiqui98}), and
the shape of the psf is similar to the shape of the cluster radial profile:  in
fact, the MECS psf is parameterized as the sum of a gaussian and a King profile.
Therefore the spectra of the different annuli need to be fitted simultaneously,
taking into account the energy-dependent overlap of the response for the
different regions.  We have taken this overlap into account in the response
matrix used for the spectral fitting.  Briefly, we took the vignetting and
strongback obscuration into account, and lacking more information we assumed the
point-spread function to be constant over the field of view.  This last
assumption is strictly true only for the central 10\arcmin, but since our
extraction radii are relatively large as compared to the spatial resolution of
the instruments no large errors are made.  The on-axis psf of the LECS and MECS
were derived from the latest (January 1999) calibration release (see
\cite{kaastra99} for further details).

Since the spatial resolution of the ROSAT/PSPC and EUVE/DS detectors is better
than that of the BeppoSAX instruments, we ignored the response overlap between
annuli in those cases.  Our final data set consists of 26 spectra from 8
different annuli obtained by 6 different instruments.  Spectral analysis was
done with version 2.0 of the SPEX package (\cite{kaastra96}).

\section{Spectral analysis}

The initial spectral model consists of a thermal plasma in collisional
ionisation equilibrium (the Mewe-Kaastra-Liedahl model) for each annulus.  We
initially assumed the cluster to be isothermal.  For the inner regions the
possible effects of resonance scattering have also been taken into account for
the iron K$\alpha$ complex.  In addition, a cooling-flow model with partial
absorption has been included for the central annulus.  The galactic column
density was fixed at $8.1\times 10^{23}$~m$^{-2}$.  Again, more details will be
presented by \cite{kaastra99}.

\placefigure{fig1}

Our best-fit model has $\chi^2=785$ for 614 degrees of freedom (dof), for a
temperature of 4.71$\pm$0.13\,keV.  This fit is formally not acceptable.  An
inspection of the fit residuals shows that there is excess flux at large radii
for both low and high energies.  This is illustrated in fig.~\ref{fig1}, showing
that there is a soft excess as compared to the thermal model in the DS data and
the LECS data below $\sim0.2$\,keV, starting from a radius of about 6\arcmin\
and increasing in relative strength up to 2--3 times the thermal count rate near
15\arcmin.  Other evidence for spectral softening at large radii is obtained
from the ratio of the Einstein IPC to ROSAT PSPC fluxes as derived from archival
data.  This ratio is constant from 2--12\arcmin, but starts decreasing beyond
that radius.

\placefigure{fig2}

Also, there is a hard excess shown by the MECS data above 7--8\,keV
(fig.~\ref{fig2}), starting at slightly larger radii than the soft excess and
also with increasing relative strength up to 2 times the thermal count rate.
Between 9--24\arcmin, the total 8--10\,keV count rate is 5.4$\pm$0.6~counts/ks,
while the best-fit thermal model predicts only 3.4~counts/ks.  For comparison,
the subtracted background is 38.2~counts/ks, of which 33.4~counts/ks can be
attributed to the particle background.  The hard excess is consistent with the
PDS data:  the observed 17--100\,keV PDS count rate is 0.090$\pm$0.024\,c/s,
while the thermal model predicts only 0.038\,c/s.  The uncertainties on the PDS
data are not sufficient to prove by themselve the existence of the hard X-ray
tail in view of the systematic uncertainties in the PDS background subtraction.
Based upon the source distribution in the HEAO-1 A4 all sky survey, the
probability to find an extragalactic hard X-ray source of this strength in the
PDS-FOV around A2199 is only $\sim$10\,\%.  From a map of the observed hardness
ratio as obtained by the MECS instrument we can conclude that the hard excess is
not due to a few discrete point sources in the field of view, but is distributed
over all azimuthal angles.

The hard tail is the {\it dominant} component outside radii of about 12\arcmin\
at energies above 8--10\,keV.  Could the hard tail have been observed before
with collimated instruments like EXOSAT or GINGA?  The answer is:  probably not.
For example, in the 8--10\,keV band it constitutes only a few percent of the
total flux.  It is only the combination of high sensitivity in the hard X-ray
band (the PDS instrument) and the spatially resolved spectroscopy of the MECS
that this component is detected.  Without spatially resolved spectroscopy a part
of the tail would be accomodated for by a slight increase in cluster
temperature.  This is confirmed by the GINGA/SSS data of \cite{white94}, who
measure a temperature of 4.74$\pm$0.09\,keV, consistent with our present fit,
and significantly larger than the ASCA temperature of 4.17$\pm$0.11\,keV in the
3--11\arcmin\ range (\cite{mushotzky96}).

In order to have a better understanding of the hard- and soft excess emission,
we extended our spectral model with a power law component in the outer radial
zones.  The photon index of this power law was kept the same for all regions.
The best-fit value for this photon index is 1.81$\pm$0.25.  The power law
component reproduces the hard excess of the spectrum as well as a part of the
soft excess.  We did not include the 25--28\arcmin\ annulus in our spectral fit,
since only ROSAT PSPC data were available for this region.  However, the
spectral shape of these outermost PSPC data was consistent with our spectral
model in the 18--24\arcmin\ annulus, and we used the PSPC data to estimate the
luminosity in this region.  The total 0.1--100\,keV luminosity of the power law
component integrated over the cluster is $(1.30\pm0.32)\times 10^{37}$~W, to be
compared to $(6.2\pm0.5)\times 10^{37}$~W for the hot isothermal component and
$(1.11\pm0.15)\times 10^{37}$~W for the cooling flow. 

We also need an additional soft component in the 6--12\arcmin\ annuli in order
to represent the DS and LECS flux in the 0.1--0.3\,keV band.  Due to the poor
spectral resolution at low energies the shape of this additional soft component
is not well constrained; we obtained good results using a narrow line feature at
0.19$\pm$0.01\,keV, but other spectral shapes can also represent the data
reasonably.  The additional soft component has an absorption-corrected
luminosity of $(1.2\pm0.3)\times 10^{36}$~W in the 0.1--0.3\,keV band, with
approximately equal luminosity in the 6--9\arcmin\ and 9--12\arcmin\ annuli.
This luminosity is 25\,\% larger than the luminosity of the thermal component in
the same annuli.  In this 6--12\arcmin\ region and 0.1--0.3\,keV spectral band
the subtracted LECS background is smaller than 15\,\% of the cluster signal,
while the large scale variations of the background in this region of the sky are
less than 10\,\% of this background.

The $\chi^2$ value for our fit is 728 for 604 degrees of freedom, an
improvement of 57 at the cost of only 10 additional parameters.  Using an F-test
we find that the excess emission is significant at the $1.8\times 10^{-6}$
confidence level.  The soft component in the 6--12\arcmin\ annuli alone is
significant at the $9.2\times 10^{-5}$ confidence level, and the power law
component alone is significant at the 0.0024 confidence level.

The best-fit temperature for the isothermal component of 4.58$\pm$0.11\,keV is
lower than the value we obtained before in our fit without the power law
component.

\section{Interpretation}

Our analysis has shown that there is both a soft and hard excess in the X-ray
spectrum of A\,2199.  Most of this excess emission can be explained by a power
law component with photon index $\approx$1.8 that dominates the cluster
luminosity outside a radius of 12\arcmin\ (640~kpc).  If the photon index is
higher in the 6--12\arcmin\ region, this could explain a part of the additional
soft excess found in that region.  Otherwise this additional component is
distinct from the power law.  Its spectral shape is not well constrained,
however its luminosity in the soft 0.1--0.3\,keV band is larger than the thermal
luminosity in the same energy band.

It is not likely that the hard excess has a thermal origin.  In that case the
temperature in the outer regions must be larger than 10\,keV (\cite{kaastra98});
this would imply that the average energy of the iron K$\alpha$ line complex
should {\it increase} by more than 100~eV from the center of the cluster towards
the edge; however the observed centroid of the K$\alpha$ blend {\it decreases}
by 60$\pm$70~eV from the center towards the 9--15\arcmin\ region (consistent
with temperatures between 1--5\,keV), hence a large temperature increase can be
excluded.  Moreover, in many nearby clusters a temperature decrease by a factor
of $\sim2$ from the center out to 6 core radii is observed
(\cite{markevitch98}).  If such a gradient would also be present in A\,2199, the
hard excess would be even stronger.

The hard excess alone can be explained by nonthermal Bremsstrahlung from a
suprathermal tail in the electron distribution (\cite{kaastra98}); however it is
not obvious how to explain the large luminosity associated with it.  Also,
although the soft excess alone can be explained by thermal radiation from very
cool gas, it requires an unrealistically large amount of rapidly cooling gas
(\cite{mittaz98}).

Given the single power law fit to both the soft and hard excess, it is more
natural to explain the observed excess by the IC emission process proposed by
\cite{sarazin98}.  Such a model was successfully applied to the soft and hard
excess in Coma (\cite{lieu99}), although it may have problems with Virgo
(\cite{reynolds99}).

Currently the best-fit single power-law index of 1.81 agrees with that of Coma
(\cite{lieu99}); both implying a number index of the relativistic electrons of
$\sim$2.62, consistent with the cosmic ray index.  A consequence of the large
pressure ratio of cosmic ray protons to electrons is that when the IC luminosity
is comparable to that of the thermal X-rays, the cosmic rays are in approximate
equipartition with the virialized gas (\cite{lieu99}).  The breakdown of
luminosity values given above suggest that A2199, like Coma, is in such a limit.
The radially rising relative importance of the IC emission, a phenomenon which
has not yet been established for Coma but was found to be present in another
cluster (A1795; \cite{mittaz98}), also follows naturally from the IC model
(\cite{sarazin98}) as a density scaling effect.  A major unresolved puzzle,
however, concerns the electrons responsible for the hard excess, as they have an
energy of $\sim$4 GeV and a resulting IC lifetime of $\sim 3\times 10^8$ years.
These electrons have to be replenished by a continuous acceleration process.

A copious source of relativistic electrons in this energy range is available,
without the need for an unrealistically high cosmic ray pressure, via the decay
of pions produced by proton-proton collisions between intracluster cosmic rays
and gas.  While in the case of a restricted injection epoch the secondaries are
not generated rapidly enough to compete against synchrotron and inverse-Compton
losses, this difficulty no longer exists if the cosmic rays have been
continuously accelerated by, e.g., intracluster shocks associated with an
on-going merger process or long-duration activity by the central radio galaxy.
Moreover the power spectral index also falls within the range of observed
values.  An outline of the model may be found in \cite{blasi99}, while detailed
development of it is currently in progress.

Another important consequence of the present data is that in the outer parts of
the cluster the contribution from the thermal component is smaller than
previously thought.  We have fitted a $\beta$-model to the emission measures of
the thermal component, and find a best fit for $\beta=0.78\pm 0.15$, a core
radius of 3.7\arcmin$\pm$0.9\arcmin\ and a central hydrogen density of
5.9$\times 10^3$~m$^{-3}$ (excluding the cooling flow contribution).  This
should be compared to e.g.  the fit of the ROSAT PSPC data by \cite{siddiqui98},
who find $\beta=0.62\pm 0.05$ with a core radius of 2.3\arcmin$\pm$0.8\arcmin.
The larger value for $\beta$ is caused by the lower thermal contribution at
large radii, the larger error bars are due to the intrinsic uncertainty in the
precise correction for the non-thermal flux.  There are important consequences
for the mass distribution within the cluster:  our parameters yield within
1.8\,Mpc (34\arcmin) a total gas mass of 7.9$\times 10^{13}$~M$_\odot$ and a
total gravitational mass of 6.8$\times 10^{14}$~M$_\odot$, a 20\,\% lower and
35\,\% higher than the mass obtained from the PSPC data of \cite{siddiqui98},
respectively.  Therefore the gas fraction is 40\,\% lower than inferred from
previous data.

\section{Conclusions}

The detection of a soft and hard X-ray excess in A\,2199 has important
consequences for the study of this and other clusters.  We demonstrated how the
excess may be explained as an IC effect, and how further investigations of this
phenomenon are vital towards understanding of cluster evolution - the history of
particle acceleration and interplay between thermal and non-thermal components
are now revealed to be much richer than previously thought.

\acknowledgments

SRON is supported financially by NWO, the Netherlands foundation for Scientific
Research.  We thank the referee for several constructive comments and
suggestions that helped us in clarifying the presentation of our results.  We
thank S.  Molendi for providing us with necessary calibration information.

\clearpage

\figcaption[fig1.eps]
{Observed count rate (symbols with errorbars) and thermal model
(histograms) for the DS and LECS (0.1--0.2\,keV) data.
\label{fig1}}

\figcaption[fig2.eps]
{Upper panel: ratio of the observed MECS count rate in the  8--10\,keV
band to the best-fit thermal model, showing the presence of the hard excess.
Lower panel: observed 8--10\,keV intensity (filled circles with error bars,
the two innermost annuli are out of scale),
the thermal model (dashed line), and the subtracted background
(triangles with error bars). The particle contribution to the 
background is shown separately as a dotted histogram.
\label{fig2}}

\clearpage

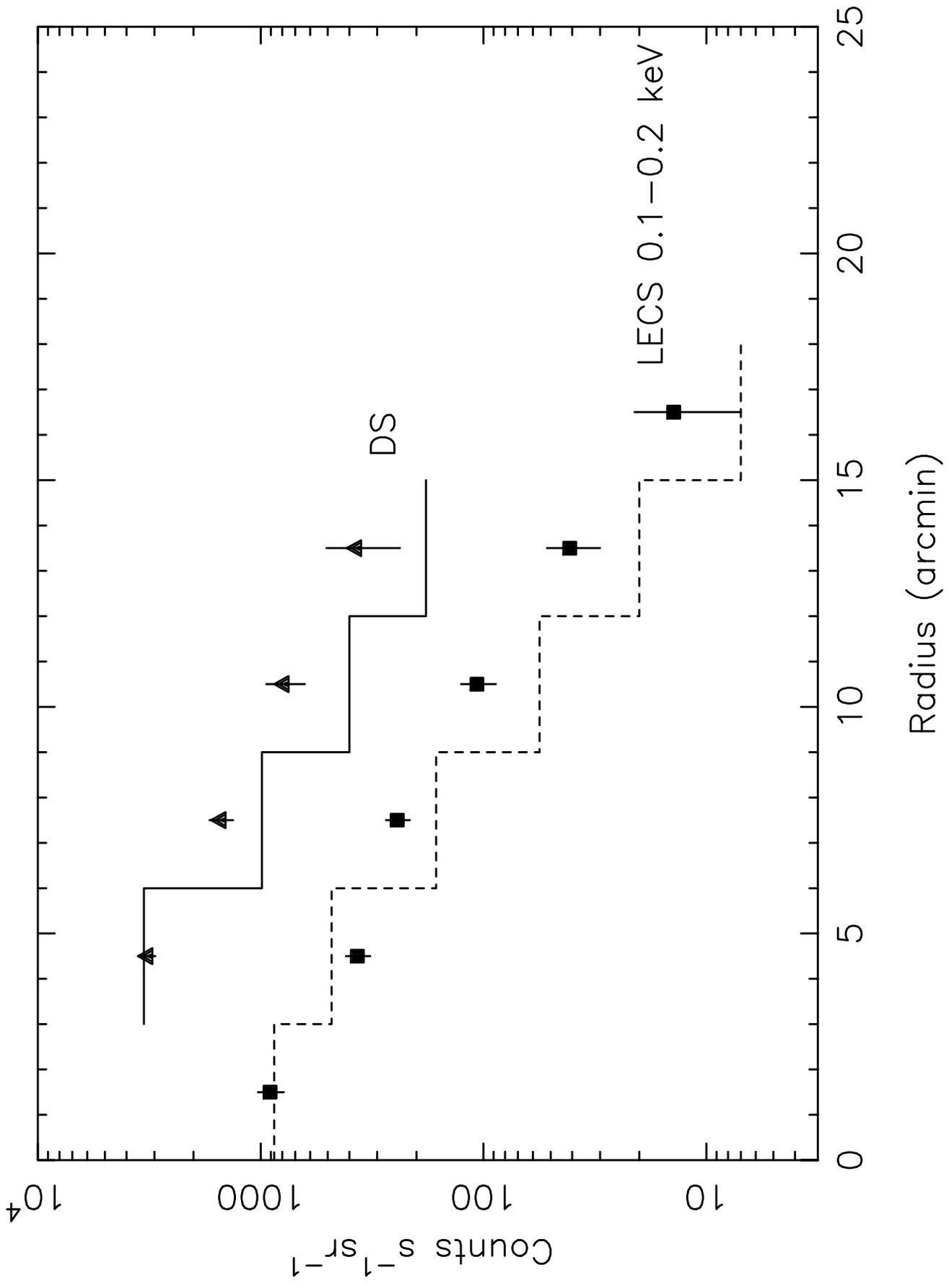
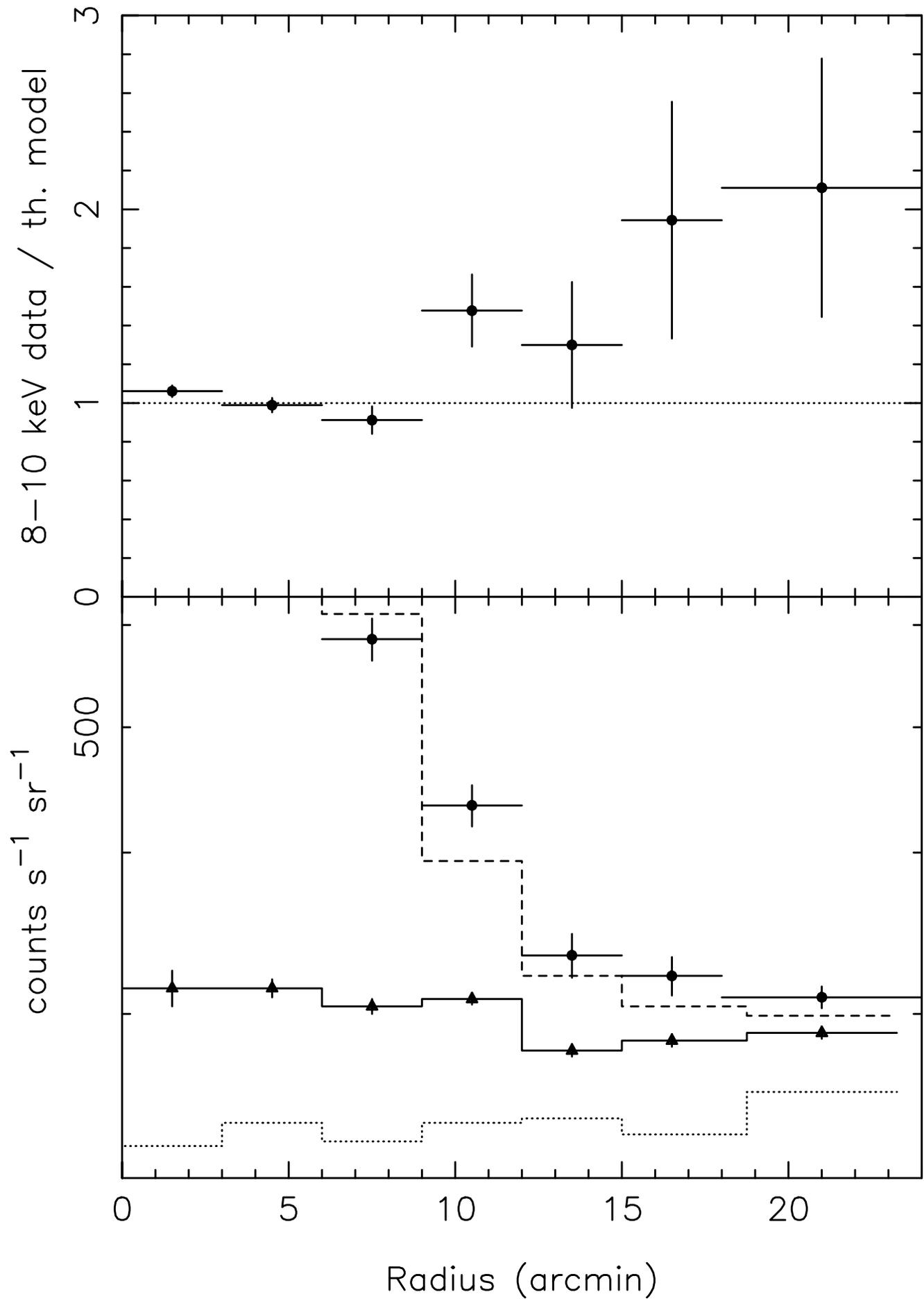
\end{document}